# Spatial regression-based transfer learning for prediction problems

Daisuke Murakami[1,2*], Mami Kajita[1], Seiji Kajita[1]

[1] Singular Perturbations Co. Ltd., 1-5-6 Risona Kudan Building, Kudanshita, Chiyoda, Tokyo, 102-0074, Japan

[2] Department of Statistical Data Science, Institute of Statistical Mathematics, 10-3 Midori-cho, Tachikawa, Tokyo, 190-8562, Japan

* Correspondence: dmuraka@ism.ac.jp

Abstract: Although spatial prediction is widely used for urban and environmental monitoring, its accuracy is often unsatisfactory if only a small number of samples are available in the study area. The objective of this study was to improve the prediction accuracy in such a case through transfer learning using larger samples obtained outside the study area. Our proposal is to pre-train latent spatial-dependent processes, which are difficult to transfer, and apply them as additional features in the subsequent transfer learning. The proposed method is designed to involve local spatial dependence and can be implemented easily. This spatial-regression-based transfer learning is expected to achieve a higher and more stable prediction accuracy than conventional learning, which does not explicitly consider local spatial dependence. The performance of the proposed method was examined using land price and crime predictions. These results suggest that the proposed method successfully improved the accuracy and stability of these spatial predictions.

1. Introduction

In the era of open data, a wide variety of spatial (and spatiotemporal) data is available for urban management, including crime prevention, environmental protection, and disaster risk management (e.g., Li et al., 2020). Because spatial data is often available only at a limited number of sample sites, spatial regression models (see Cressie, 2015; Cressie and Wikle, 2015; see Section 2) have been used to predict/interpolate missing observations. For example, Cattle et al. (2002) predicted soil contamination, Mercer et al. (2011) predicted air quality and Tsutsumi et al. (2011) predicted land price.

Spatial regression model can perform poorly if the sample size in a spatial area where we want to predict, called the target area, is small. In such a case, transfer learning (see Zhuang et al. 2020) helps improve the prediction accuracy using samples obtained in areas outside the target area, which we will call source areas. For example, crime prediction accuracy in a target area may be improved using samples from other source areas. As reviewed by Weiss et al. (2016) and Zhuang et al. (2020), there are a wide variety of approaches for transfer learning.

Transfer learning has been used for spatial prediction (e.g., Zhang et al., 2019; Mao, 2020; Iwata and Tanaka, 2022). Recent relevant studies have typically employed neural networks to associate source and target areas. Because neural networks require many observations to attain a high spatial prediction accuracy (see Seya and Shiroi, 2022), massive samples are usually assumed in the source areas. For instance, Zhang et al. (2019) used cellular traffic data at 10 min intervals with 300 million records, whereas Iwata and Tanaka (2022) used 50 100 × 100 km regions, each of which has 20 attribute values by a 1 × 1 km grid square. Such fine-grained observations are typically available if the analyzed data are remotely sensed images, traffic counts, or other sensor data.

However, such massive amounts of source data are often unavailable. For example, social survey data, such as population, income, crime counts, and the number of COVID-19 positives, are typically available only at the municipality and district levels to protect privacy. Data with high acquisition costs, such as ecological and geological data, are often available only at a limited number of sample sites. When applying transfer learning to such data, the available data size in the source areas would be considerably smaller than those assumed in existing studies. Because spatial regression tends to be more accurate than methods that ignore spatial dependence when the sample size is small (see Section 2), relying on spatial models in small-scale transfer learning tasks is sensible.

To our knowledge, the conventional spatial regression model has never been extended to transfer learning. Therefore, this study develops a spatial regression-based transfer learning method that improves spatial prediction accuracy in a target area where relatively small samples, such as tens of thousands of samples, are available in source areas. We use spatial regression models to capture local features in each area and a machine learning algorithm to balance the local features and common features across areas.

The remainder of this paper is organized as follows. Section 2 reviews the spatial prediction method and provides a motivational example. In Section 3, the proposed method is introduced. Sections 4 and 5 apply our method to land price and crime prediction to examine the performance of the proposed method. Finally, Section 6 concludes the paper.

2. Motivative example

The spatial dependence that "nearby things are more related than distant things" (Tobler, 1970) has long been recognized as one of the most important properties for accurate spatial prediction (see Cressie and While, 2015). In spatial regression, spatial dependence is modeled using a Gaussian process (GP; Williams and Rasmussen, 2006) whose covariance decays with respect to the Euclidean distance. From a Bayesian perspective, the spatial GP acts as a prior/regularizer to stabilize the model. In other words, the model is estimated considering data fitting and consistency with prior information that the data are spatially dependent. The spatial prior prevents overfitting, and the resulting prediction tends to be stable and accurate, even with small samples, as illustrated in Figure 1.

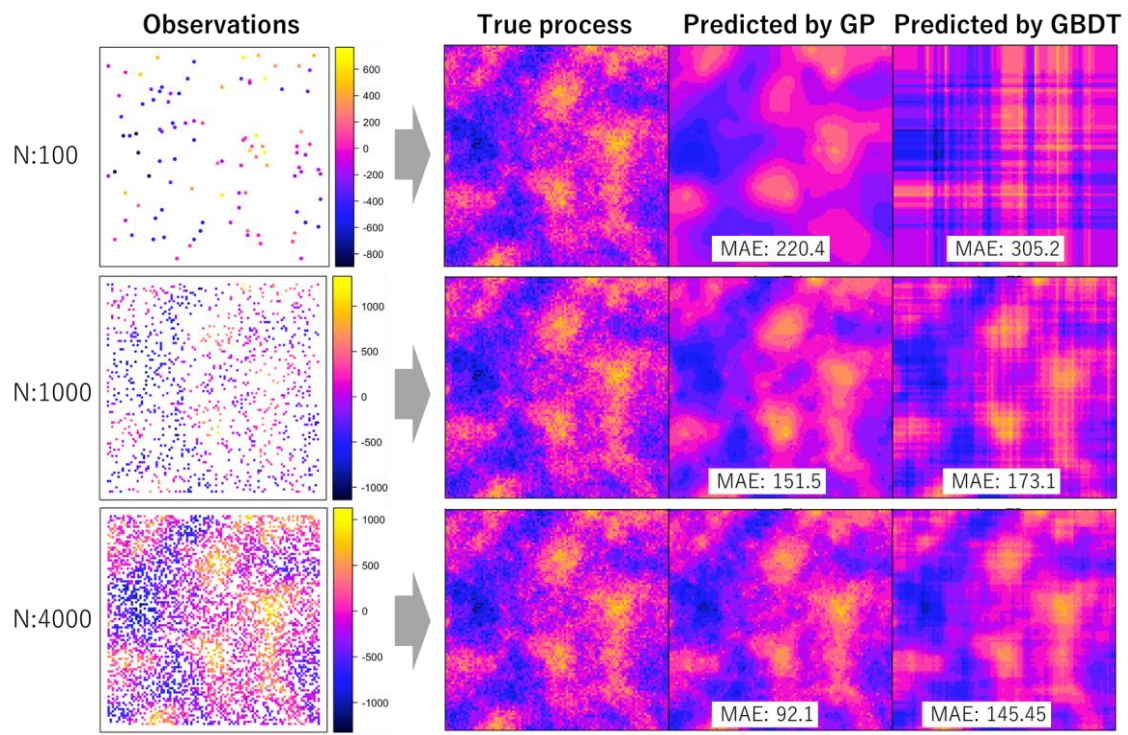

Figure 1: Toy example of spatial predictions by GP and GBDT. Here, the true process is generated on 100 × 100 grid points using a GP defined by $z_i + e_i$ where $z_i \sim N(0, \exp(-d_{ij}))$ is a spatially dependent process where $d_{ij}$ is the distance between sites and $e_i \sim N(0,1)$ is a white noise. For samples with $N \in \{100, 1000, 4000\}$, GP and GBDT are fitted, and their predictive accuracy on the unobserved grid points is evaluated using mean absolute error (MAE).

On the other hand, gradient boosting (GB; Friedman, 2001), neural network (NN), and other machine learning algorithms have also been used for spatial prediction. These techniques typically consider spatial coordinates and timestamps as input and other features (i.e., explanatory variables). These algorithms flexibly learn complex spatiotemporal patterns for large samples and attain better predictive accuracy than conventional spatial regression methods. However, because these algorithms do not explicitly consider spatial dependence, they often fail to capture local spatial patterns underlying small samples. For example, Figure 1 compares GP and gradient-boosting decision trees (GBDT) through spatial prediction. GBDT is a popular algorithm for learning data that uses an ensemble of decision trees grown to minimize a loss function, which is the squared error in our case. Because the true data are generated from GP, GBDT is less accurate than GP. However, the accuracy of the GBDT is considerably worse than that of the GP. In particular, when $N = 100$, the GBDT exhibits spurious map patterns. The naïve NN, which does not impose a spatial prior, suffers from the same problem as long as only small samples are available in the target area.

In short, if small samples are available in the target area, it would be reasonable to use GP-based spatial regression to capture local features.

3. Methodology

Our objective was to improve the spatial prediction accuracy in the target area through transfer learning from large samples in the $G$ source areas. We assumed source areas where large samples were available and a target area where only small samples were available. Considering the discussion in Section 2, we apply spatial regressions to model local features in the source and target areas, while GB is used to model common features across areas. In each region, the explained and explanatory variables were assumed to be scaled to mean zero and variance, respectively.

Specifically, let $i$ and $I(g)$ denote the indices of the sample sites in the target and $g$-th source areas, respectively. The explained variables $y_i$ and $y_{I(g)}$, where $g \in \{1, \dots, G\}$, are modeled by the following equations:

$$y_i = f_\theta(\mathbf{x}_i, z_i) + u_i, \tag{1}$$

$$y_{I(g)} = f_\theta(\mathbf{x}_{I(g)}, z_{I(g)}) + u_{I(g)}. \tag{2}$$

$\mathbf{x}_i \in \{x_{i,1}, \dots, x_{i,K}\}$ and $\mathbf{x}_{I(g)} \in \{x_{I(g),1}, \dots, x_{I(g),K}\}$ represent the $K$ explanatory variables in the target and source regions, $z_i$ and $z_{I(g)}$ denote the local spatial features in each region, and $u_i$ and $u_{I(g)}$ are the noise terms. The function $f_\theta(\cdot)$ which depends on the parameter $\theta$ is commonly used in Eqs. (1) and (2), based on the assumption that the explained variables and local features influence the explained variable in the same manner in each area. The couple of Eqs. (1) and (2) is a general formulation for multitask/transfer learning that models common features across the source and target areas. While NN, GBDT, or other functions are available for $f_\theta(\cdot)$, we use GBDT because of its flexibility and ease of tuning compared with NN. These advantages are expected to be useful for relatively small samples, as in the present case. Eq. (1) is identical to the conventional GBDT, if $z_i = 0$. In other words, we attempt to improve the modeling accuracy in the target area by introducing local feature $z_i$.

As discussed in Section 2, the GBDT can fail to capture local spatial patterns in the target area. Therefore, the local features $z_i$ and $z_{I(g)}$ are pretrained in each region using spatial GPs, such that

$$y_i = z_i + e_i, \qquad e_i \sim N(0, \sigma^2 w_i^{-1}), \tag{3}$$

$$y_{I(g)} = z_{I(g)} + e_{I(g)}, \qquad e_{I(g)} \sim N(0, \sigma^2 w_{I(g)}^{-1}), \tag{4}$$

where $w_i$ is a known weight. $z_i$ and $z_{I(g)}$ are GPs. GP is computationally demanding. Therefore, to reduce the computational cost in source areas where the sample size is large, we assume the following low-rank GPs, which can be computationally efficient (see Murakami and Griffith, 2019a):

$$z_i = \sum_{k=1}^{K} x_{i,k} b_k + \sum_{l=1}^{L} s_{i,l} \gamma_l, \qquad \gamma_l \sim N(0, \tau^2 \lambda_l^\alpha), \tag{5}$$

$$z_{I(g)} = \sum_{k=1}^{K} x_{I(g),k} b_{k(g)} + \sum_{l=1}^{L} s_{I(g),l} \gamma_{l(g)}, \qquad \gamma_{l(g)} \sim N\left(0, \tau_g^2 \lambda_{l(g)}^{\alpha_g}\right), \tag{6}$$

where $b_k$ denotes the $k$th regression coefficient. The term $\sum_{l=1}^{L} s_{i,l} \gamma_l$ is a low-rank spatial process where $[s_{1,l}, \ldots, s_{N,l}]'$ and $\lambda_l$ is the $l$-th eigenvector (or eigenfunction) and the corresponding eigenvalue of a spatial kernel matrix, which is given by a matrix whose ($I$, $j$)-th element equals $\exp(-d_{i,j}/r)$. Following Dray et al. (2006) and Murakami and Griffith (2021), the $r$ parameter was given by the maximum edge length of the minimum spanning tree connecting the sample sites. For large samples, the eigenpairs can be approximated (see Murakami and Griffith, 2019b). The spatial pattern of $\sum_{l=1}^{L} s_{i,l} \gamma_l$ depends on parameters $\tau^2$ and $\alpha$ which determine the variance and spatial scale of the process. $\{x_{I(g)}, b_{k(g)}, s_{I(g)}\}$ are defined in a manner similar to $\{x_{i,k}, b_k, s_i\}$.

The local features $z_i$ and $z_{I(g)}$ are characterized by area-specific regression coefficients $\{b_k, b_{k(1)}, \ldots, b_{k(G)}\}$ and area-specific spatial processes. Eqs. (5) and (6) were estimated using fast marginal likelihood maximization (Murakami and Griffith, 2019a). Given $\hat{b}_k$ and $\hat{\gamma}_l = E[\gamma_l | y_i, \hat{\boldsymbol{\theta}}_z]$ where $\hat{\boldsymbol{\theta}}_z$ is the estimate of $\boldsymbol{\theta}_z \in \{b_1 \ldots, b_K, \tau^2, \alpha, \sigma^2\}$, the conditional expectation of $z_i$ is $\hat{z}_i = \sum_{k=1}^{K} x_{i,k} \hat{b}_k + \sum_{l=1}^{L} s_{i,l} \hat{\gamma}_l$. Likewise, the expectation of the local feature $\hat{z}_0$ in an unobserved site is given as:

$$\hat{z}_0 = \sum_{k=1}^{K} x_{0,k} \hat{b}_k + \sum_{l=1}^{L} s_{0,l} \hat{\gamma}_l, \tag{7}$$

where $x_{0,k}$ is the $k$-th explanatory variable at the site and $s_{0,l}$ is the value of the $l$-th eigenvector (or eigenfunction) extended through the Nyström method (Drineas et al., 2005). Similarly, $\hat{z}_{I(g)} = \sum_{k=1}^{K} x_{I(g),k} \hat{b}_k + \sum_{l=1}^{L} s_{I(g),l} \hat{\gamma}_{l(g)}$ is available after estimating $\boldsymbol{\theta}_Z \in \{b_{k(1)}, \ldots, b_{k(G)}, \tau^2, \alpha, \sigma^2\}$ for each source area.

Unfortunately, it is difficult for a naïve multitask GBDT (Eqs. 1-2) to capture the local features in the target area (see Section 2); we addressed this difficulty by pre-training $\hat{z}_i, \hat{z}_{I(1)}, \ldots, \hat{z}_{I(G)}$ using spatial regressions. We predicted the unobserved explained variable $y_0$ in

the target region as follows:

(1) Standardize the explained variables and explanatory variables with mean zero and variance one.
(2) In the target area, fit the spatial model Eq. (3) whose $z_i$ is given by Eq. (5). Then predict the local features $\hat{z}_i, \hat{z}_0$ at observed and unobserved sites using the expectation of Eq. (5) and Eq. (7), respectively:
(3) In each source area, fit the spatial models Eq. (4) whose $z_{I(g)}$ is given by Eq. (5). Subsequently, predict the local features $\hat{z}_{I(1)}, \ldots, \hat{z}_{I(G)}$ in each area using the expectation of Eq. (6).
(4) Train the multitask GDBT model (Eqs. 1 and 2) using $\hat{z}_i, \hat{z}_{I(1)}, \ldots, \hat{z}_{I(G)}$ to determine function $\hat{f}_\theta(\cdot)$.
(5) Predict $y_0$ by using the trained GBDT model $\hat{f}_\theta(\cdot)$ with $\mathbf{x}_0 \in \{x_{0,1}, \ldots, x_{0,K}\}$ and $\hat{z}_0$.

The spatial regression model used in steps (2) and (3) can be substituted with other spatial models, including the exact GP (i.e., kriging), spatial autoregressive models (see LeSage and Pace, 2005), and geographically weighted regression (Brunsdon et al., 1998).

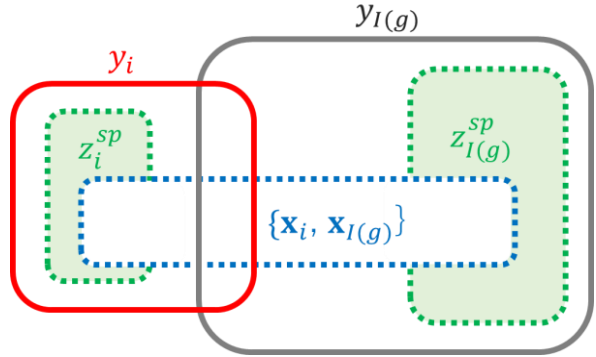

Figure 2: Venn diagram of $y_i$, $y_{I(g)}$, and other variables assumed in this study. $z_i^{sp} = \sum_{l=1}^{L} s_{i,l} \gamma_l$ and $z_{I(g)}^{sp} = \sum_{l=1}^{L} s_{I(g),l} \gamma_{l(g)}$. The size of the red area denotes the total variations of $y_i$, and the interior represents the variance composition explained by explanatory variables (blue area) and spatial dependent features (green areas). The gray area is similarly defined for $y_{I(g)}$. A larger overlap between two areas/variables means a stronger dependency between the two variables. For example, the red and grey areas overlap because $y_i$ and $y_{I(g)}$ are assumed to have commonality, e.g., explained by the explained variables, whereas the two green areas are non-overlapping because spatially dependent features in each region are assumed mutually independent.

Figure 2 shows the Venn diagram. $y_i$ and $y_{I(g)}$ are assumed to depend on the common explanatory variables/features $\mathbf{x}_i$ and $\mathbf{x}_{I(g)}$. Such common features are helpful for improving the transfer learning accuracy. However, spatial patterns in each region are usually independent of each other, as illustrated in the figure (e.g., even if there is a crime hotspot in the northern part of a study area, the same does not necessarily hold in another study area), and is difficult to transfer. Therefore, in this study, the spatially dependent patterns in each region (the green region in the figure) were pre-trained and embedded in the local features $\hat{z}_i, \hat{z}_{I(1)}, \ldots, \hat{z}_{I(G)}$ in step (3). Subsequently, the influences of these features are evaluated through multitask learning in step (4). Thus, our proposed method learns the local spatial features in each region (green region) and the common features (blue region) across regions.

Although this approach is simple and easy to implement, its predictive accuracy outperforms the conventional GBDT, while it is more stable than the conventional spatial regression model, as demonstrated in the subsequent sections.

4. Experiment 1: Land price modeling

4.1. Setting

This section examines the predictive accuracy of the proposed model with existing methods through Monte Carlo experiments using residential land price data officially assessed in 2015 in the Tokyo and Osaka metropolitan areas, which are the largest urban areas in Japan (see Figure 2). The sample sizes were 10,540 in the Tokyo region and 6,164 in the Osaka region.

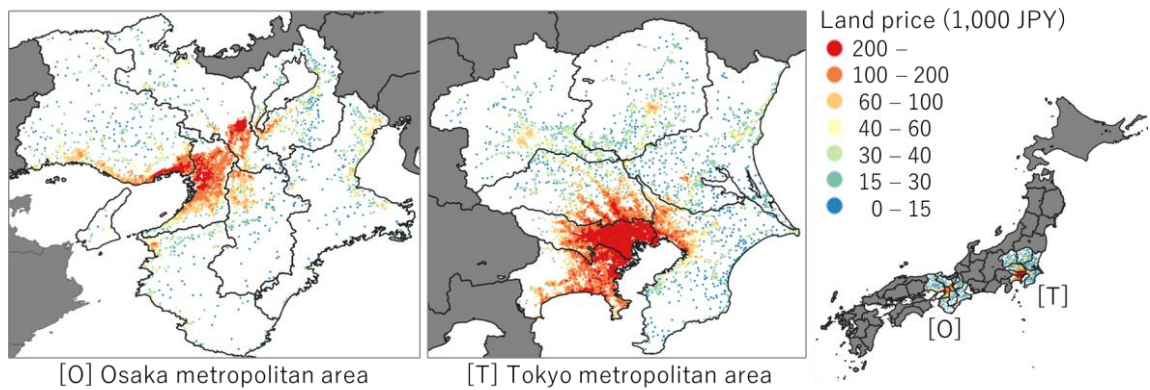

Figure 2: Residential land price in the Tokyo and Osaka metropolitan areas in 2015 and the prefectures (right). As shown in this figure, the metropolitan areas consist of seven prefectures.

We perform transfer learning from the Tokyo metropolitan area, where all the samples are assumed to be available, to predict land prices in the Osaka metropolitan area, where only $N_{obs}$ samples are assumed to be available. In Osaka, the $N_{obs}$ entries were sampled randomly for training, and the remaining $N - N_{obs}$ samples were used to evaluate the predictive accuracy. We set $N_{obs} \in \{20, 50, 100, 500, 1000\}$. For each $N_{obs}$, spatial prediction was performed 200 times.

The proposed method is compared with the conventional linear regression model (LM) and spatial LM (SPLM), a linear regression with residual spatial dependence consisting of Eqs. (3) and (5), respectively. LM and SPLM consider only the $N_{obs}$ samples. The proposed method is also compared with two transfer learning methods, including a generalized linear model-based transfer learning technique (GLMtrans) of Tian and Feng (2022) and the conventional GBDT, ignoring the local features (i.e., Eqs.1 and 2 without $z_i, Z_{i(1)} \ldots, Z_{i(G)}$). We also estimated a non-transfer GBDT by considering only the $N_{obs}$ samples (GBDT$_{loc}$). GLMtrans assumes linear regression models in the source and target regions and balances each model through ridge regularization. In the GBDT, we used an ensemble of 3,000 decision trees with a maximum interaction depth of 3 to minimize the squared error. For training, half of the samples, which were randomly sampled, were used, and the other half was used for evaluating the out-of-sample squared error for learning. The prediction accuracy of each method was evaluated using the root mean square error (RMSE).

For each model, the explained variables are the distance to the nearest railway station [km], distance to the nearest bus station [km], distance to the nearest prefectural capital [km], and elevation [m]. These variables were acquired from the National Land Numerical Information download service (https://nlftp.mlit.go.jp/ksj/) provided by the Ministry of Land, Infrastructure, Transport, and Tourism (MLIT), Japan. Our preliminary analysis suggested that considering longitude and latitude in the GBDT and GLMtrans did not improve the prediction accuracy in the target area, probably because the target area does not overlap with other areas. The spatial coordinates were not included as explanatory variables.

4.2. Result

Figure 3 plots the mean RMSE values over the 200 simulations in each case. Although the flexibility of GBDT is widely recognized, GBDT$_{loc}$ did not perform well due to the small samples; the result is consistent with Figure 1. SPLM and our method outperform the other methods when $N_{obs} \geq 50$. The importance of considering spatial dependence is confirmed. The proposed method tends to have smaller RMSEs than SPLM across cases; in particular, their difference is considerable when $N_{obs}$ is small.

GLMtrans and GBDT did not improve their accuracy, even when $N_{obs}$ were large. This might be because they emphasize common features across areas while failing to capture local model features. Notably, the only difference between the GBDT and our method is that the latter considers pre-trained local features. The finding that the local feature significantly improves transfer learning accuracy is interesting.

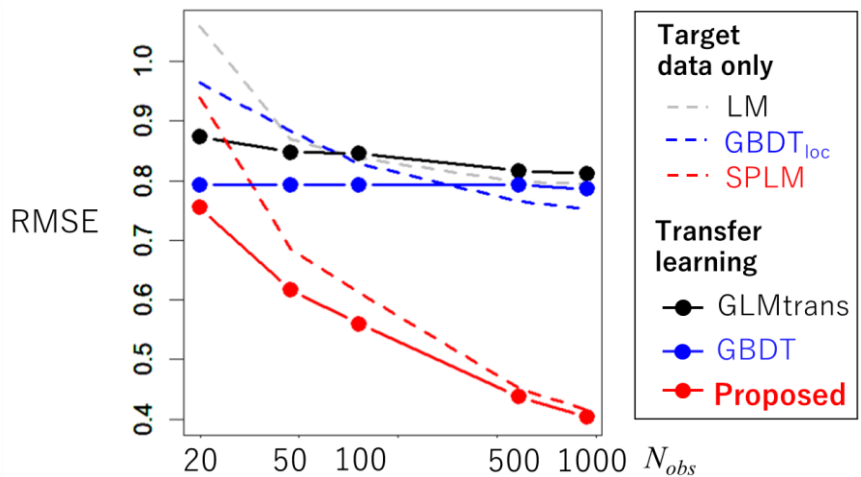

Figure 3: Mean RMSEs for the land price prediction. The x-axis is the sample size in the target area ($N_{obs}$) in the log-scale.

Figure 4 displays the boxplots of the RMSE values when $N_{obs} \in \{20, 50\}$. For small samples, LM and SPLM had large variations in their RMSEs, suggesting their instability. The RMSE variation for the proposed method was considerably smaller than those of the two methods. It was verified that our transfer learning stabilized the spatial prediction.

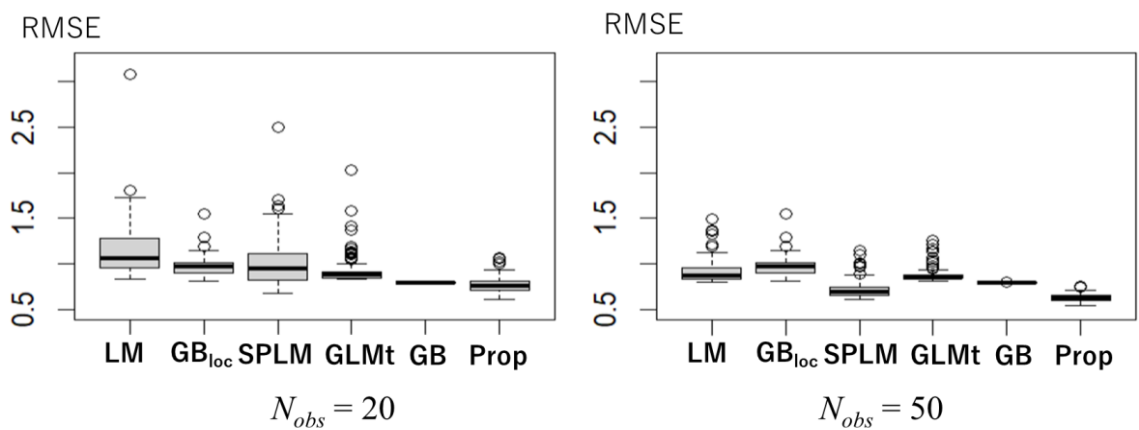

Figure 4: Boxplots of the RMSE values when $N_{obs} \in \{20, 50\}$ (GB$_{loc}$ = GBDT$_{loc}$; GLMt = GLMtrans; GB = GBDT; Prop = Proposed method).

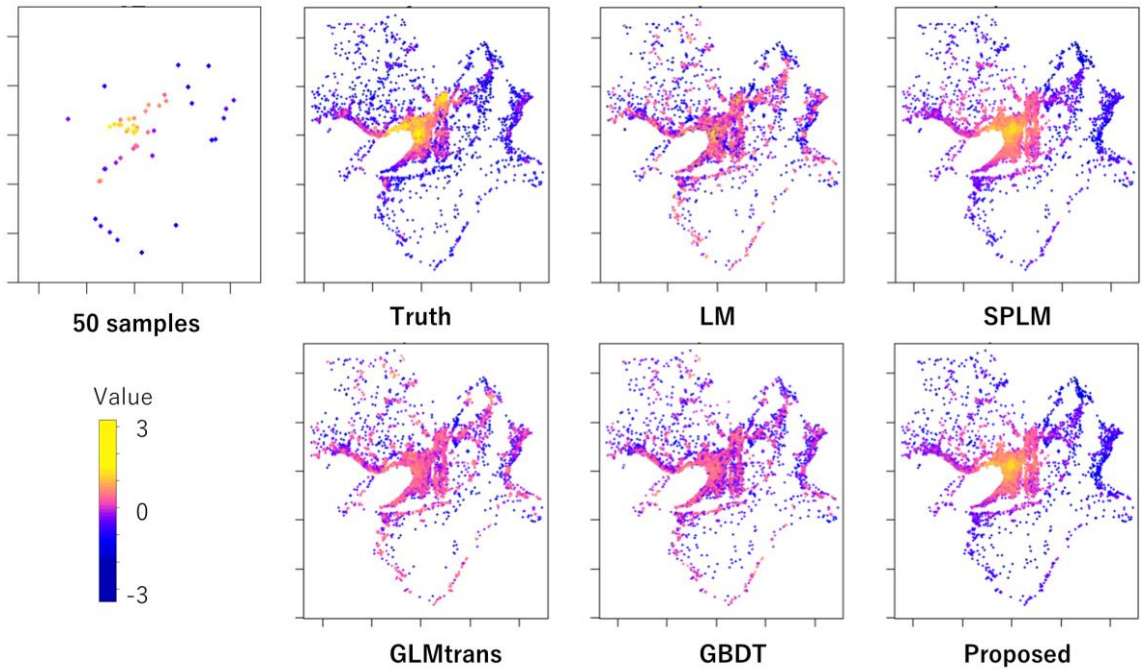

Figure 5: An example of spatial interpolation result when $N_{obs}$=50. The 50 samples used for model estimation (50 samples), the true land price distribution (Truth), and predicted land prices are plotted. Note: land price values are standardized.

Figure 5 shows spatial prediction results obtained in an iteration with $N_{obs} = 50$. SPLM and the proposed method have similar map patterns with the truth. The other methods ignoring spatial dependence fail to capture the high land price in the center. Modeling spatial dependence is again found to be necessary to estimate the map pattern of the land prices accurately.

To compare these results in more detail, Figure 6 shows plots where the *X*- and *Y*-axes are the true and predicted land price values. The predictive accuracy of SPLM is unsatisfactory in the low-price range. This may be because low-price samples outside the center are scarce, as shown in Figure 6 (top left). GBDT has a better fit than SPLM in the low-price range (see the bottom middle of Figure 6) owing to transfer learning. However, the GBDT, which ignores spatial dependence, is less accurate overall. In contrast, our GBDT extension, which considers spatial dependence, accurately predicts land prices across value ranges. Our method is a reasonable choice for the spatial transfer learning of regional data.

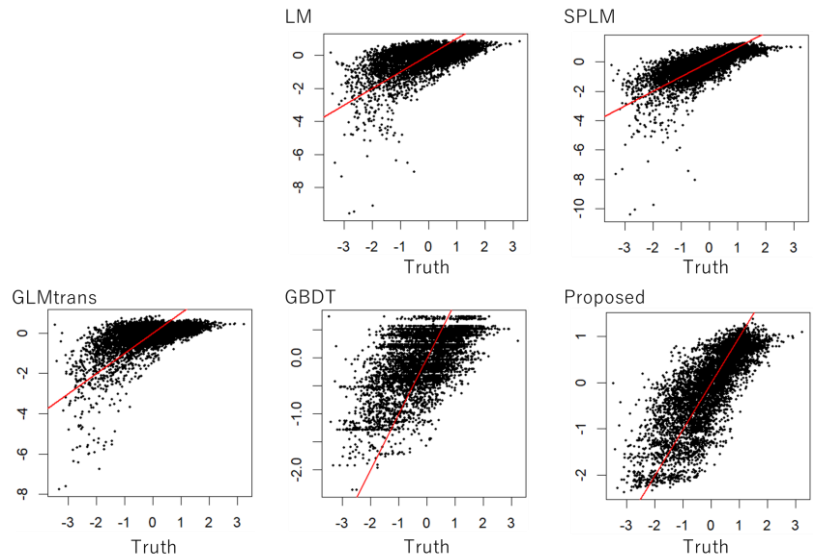

Figure 6: Comparison of the true and predicted land price values

## 5. Experiment 2: Crime prediction

### 5.1. Setting

This section applies the proposed method for the one-month-ahead prediction of crime counts by the district in the USA. The datasets used were those of San Francisco (SF) (source: Crime Dashboard: https://www.sanfranciscopolice.org/stay-safe/crime-data/crime-dashboard), Chicago (source: Chicago Data Portal: https://data.cityofchicago.org/), Denver (source: Denver Open Data Catalog: https://denvergov.org/opendata), and Boston (Crime Statistics, Mass.gov: https://www.mass.gov/crime-statistics). The number of districts in SF is 196, whereas those in Chicago, Denver, and Boston are 796, 144, and 180, respectively. We examined the benchmarks of the proposed method by setting SF and Chicago as the target areas. We assumed that district crime counts were available only one month ago in a target area (SF or Chicago), while those in the past 12 months were available in the other three cities, which are the source areas. We examined whether the prediction accuracy was improved by considering the crime counts in other cities. The types of predicted crime counts were larceny, robbery, and burglary (see Figure 7 for the case of SF). Predictions were performed 12 times for each while varying the target month from January to December 2018.

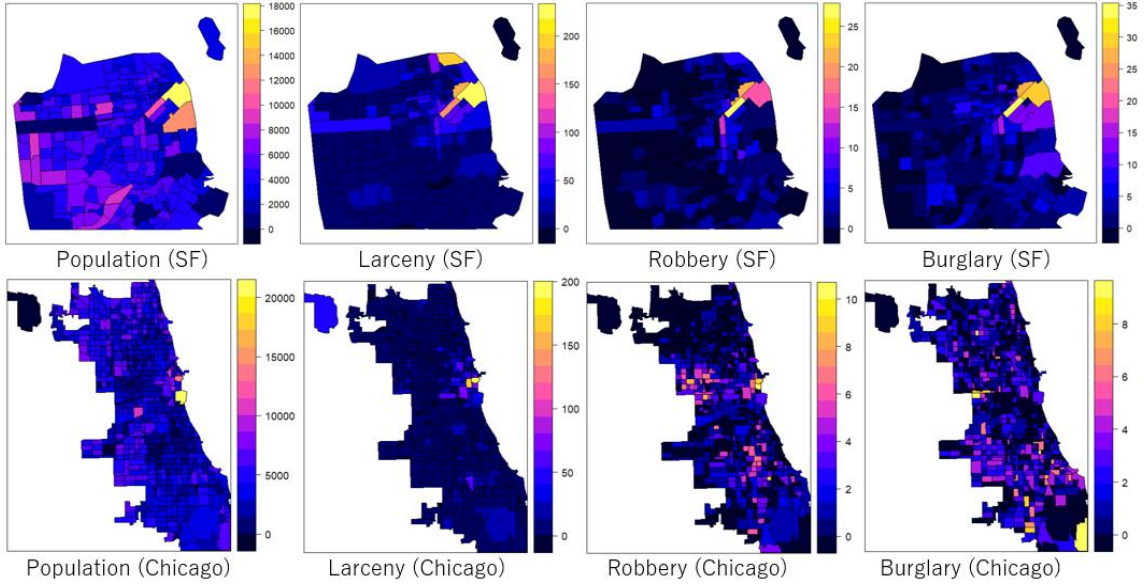

Figure 7: Crime counts by district in San Francisco (top) and Chicago (bottom) in July 2018.

To model crime counts, it is necessary to assume a probability distribution for counts, such as a Poisson distribution, rather than a Gaussian distribution, which we have assumed so far. To apply our method to crime data, we transformed our Gaussian model to an approximate Poisson model following Murakami and Matsui (2022). They showed that a spatial (over-dispersed) Poisson regression model is approximated by a spatial linear regression model with the explained variable $y_i = \log(\tilde{y}_i + 0.5) - \frac{1+0.5q}{\tilde{y}_i+0.5}$ and sample weight $w_i = \tilde{y}_i + 0.5$ where $\tilde{y}_i$ is the explained count variable in the $i$-th district, and $q$ is the ratio of zero count in the target region. We specify $y_i$ and $w_i$ as explained and $y_{I(g)}$ and $w_{I(G)}$ in the same way to assume an over-dispersed Poisson distribution for crime counts. Despite its simplicity, their approximation mitigates an identification problem (Correia et al., 2019), and the resulting accuracy is even better than that of the basic spatial Poisson regression for small samples, as in our case (see Murakami and Matsui, 2022).

The proposed method is compared to the Poisson regression (PM) and spatial Poisson regression model (SPPM), which apply the Poisson approximation to SPLM, GLMtrans, and GBDT, which minimize the Poisson loss. PM and SPPM consider only the samples in the target area, whereas GLMtrans and GBDT also consider larger samples in the source area. In addition, the GBDT considering only a small sample in the target area denoted as $GBDT_{loc}$, was also compared.

The explanatory variables are two education-related variables (the ratio of residents who have no educational attainment; the ratio of residents who have a bachelor's degree), two income-related variables (median income; the ratio of people whose income is below the poverty threshold), nighttime population, longitude, and latitude.

5.2. Result

Figure 8 summarizes the predictive accuracy measured by the RMSE. SPPM outperformed PM in all cases, confirming the importance of modeling spatial dependence in crime prediction. It was also verified that GLMtrans and GBDT outperformed PM, indicating that transfer learning improved the modeling accuracy. Probably because the sample sizes in the target areas are not too small (the numbers of districts in SF and Chicago are 196 and 796, respectively), the accuracy of $GBDT_{loc}$ is comparable to that of GBDT. Our method tends to have smaller RMSEs across San Francisco and Chicago cases than these alternatives. The improvement in accuracy achieved by our method considering both local spatial features and common features was verified.

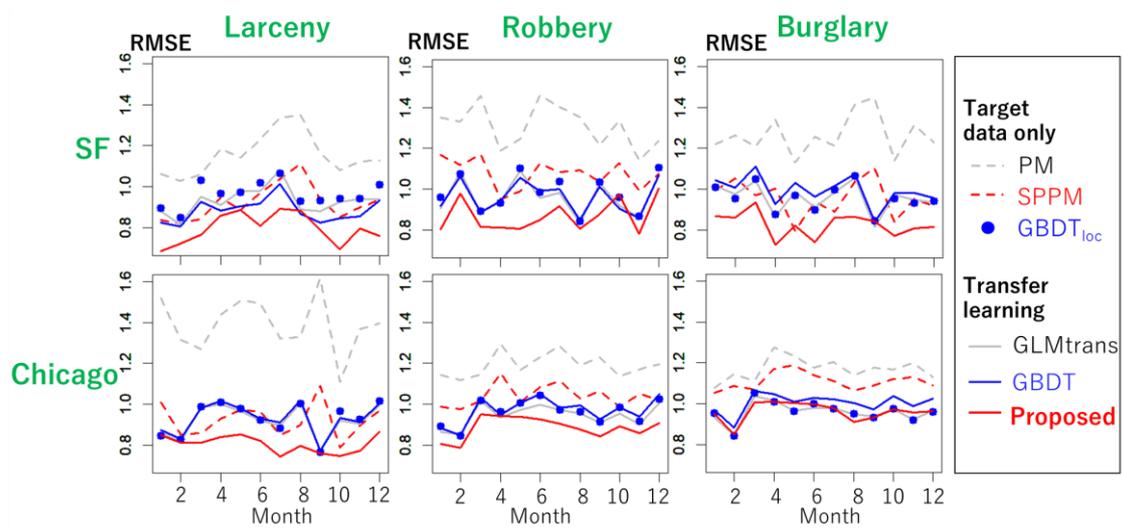

Figure 8: RMSE of the one-month-ahead crime prediction by month, city, and crime type.

Figure 9, 10, and 11 plot the observed and predicted counts of larcenies, robberies, and burglaries in SF. The northeastern area, the central area, is the hotspot for all crimes. Each model successfully identifies a hotspot. However, the risk maps estimated from PM and SPPM tended to have larger variations than the truth in suburban areas. This tendency is prominent in PM, which ignores spatial dependence. In contrast, the risk maps produced from GLMtrans and GBDT, which consider crime counts in other cities, tend to have smaller variations than the truth. Specifically, GLMtrans underestimated crime counts in the northeastern, central area. The GBDT tends to overestimate crime counts in other areas. These results suggest the need for an adjustment to consider local spatial patterns, for example, using spatial regression. The proposed method, transfer learning with such a local adjustment, successfully estimated larger crime counts in the northeastern area and smaller crime counts in other areas. As a result, the RMSE values were small relative to the others. The accuracy of the proposed method was verified.

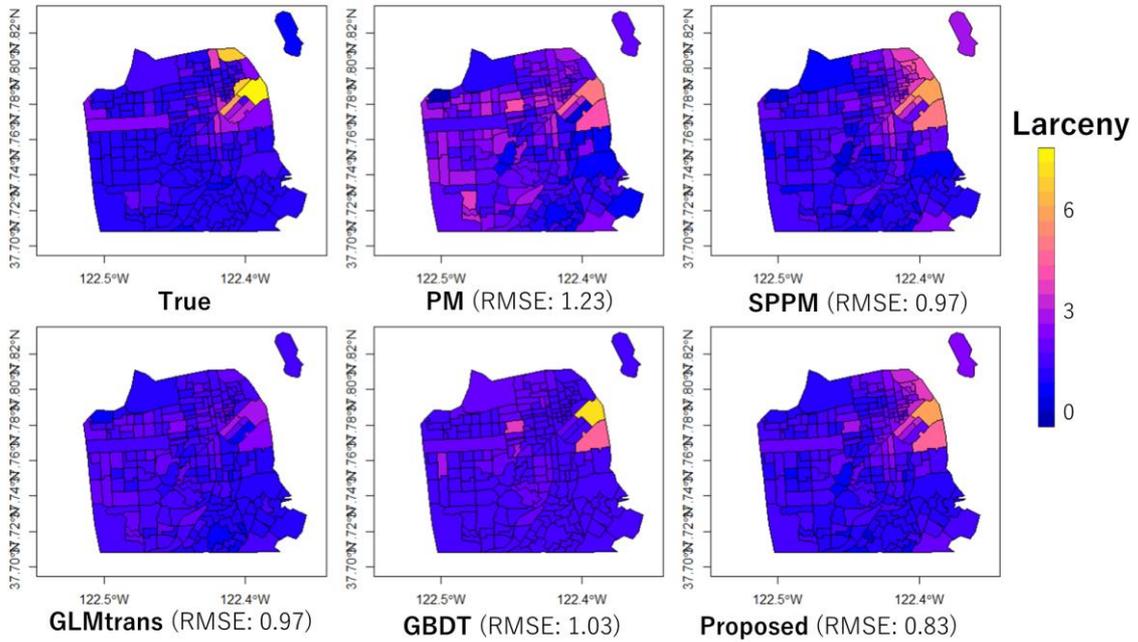

Figure 9: Observed and predicted counts of larceny (standardized) in San Francisco in June 2018.

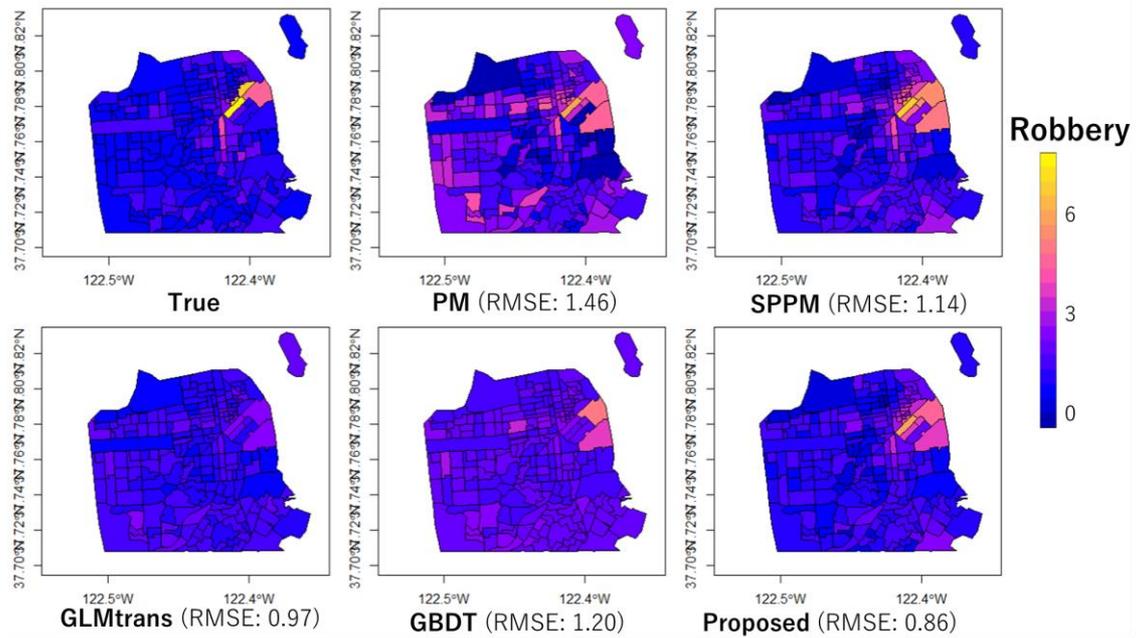

Figure 10: Observed and predicted robbery (standardized) counts in San Francisco in June 2018.

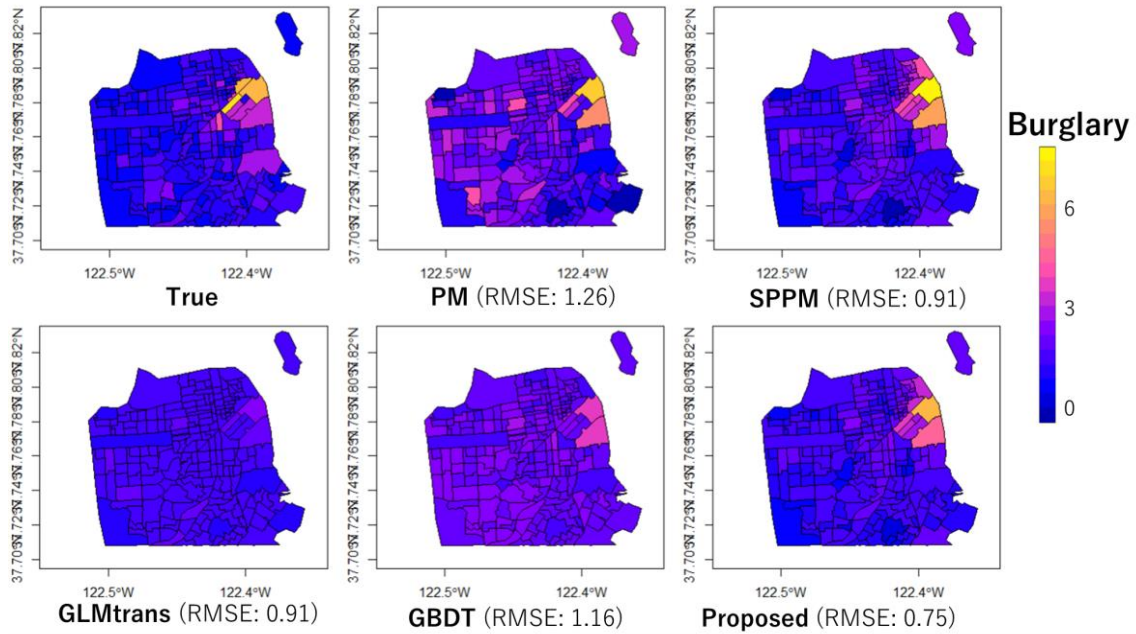

Figure 11: Observed and predicted counts of burglary (standardized) in San Francisco in June 2018.

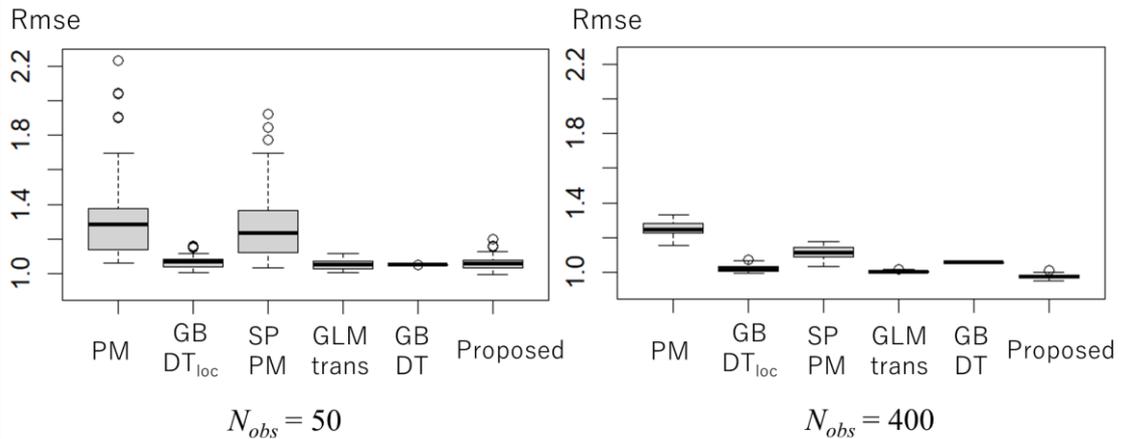

Figure 12: Boxplots of the RMSEs of the predicted values when the target sample size $N_{obs}$ is small. The target area is Chicago. Robbery counts in randomly selected $N_{obs}$ districts are assumed to be available in the target area, and a one-month prediction is performed using the data (and data in the other three cities in 12 months in case of transfer learning) 200 times.

As an additional experiment, robbery counts were assumed to be available only in randomly selected $N_{obs}$ districts in Chicago to compare the predictive accuracy for small samples. Then, in Chicago, the same one-month-ahead prediction of robbery counts is iterated 200 times for each $N_{obs} \in \{50, 400\}$. Figure 12 shows the boxplots of the resulting 200 RMSEs. The RMSE values of PM, SPPM, and $GBDG_{loc}$, which only consider Chicago data, have larger variations owing to the small samples. GLMtrans, GBDT, and our method, which are transfer learning methods, tend to have smaller and more stable RMSE values, suggesting the accuracy of these methods. In particular, the proposed method exhibited the minimum median RMSE when $N_{obs} = 400$. However, the proposed method tended to have larger RMSE variations relative to GLMtrans and GBDT when $N_{obs} = 50$, as in the previous section, owing to the inclusion of the local feature. Improving the stability of the proposed method when $N_{obs}$ are very small is an important future task.

6. Concluding remarks

This study developed a spatial regression-based transfer learning method that considers spatial dependence and common features across source and target areas and verified its predictive accuracy.

We performed transfer learning by estimating a spatial regression model and learning the GBDT. Because both methods are implemented in packages in R, a free statistical software or other software, the proposed method is easily implemented. In addition, it can easily be extended. For example, replacing the assumed spatial features with spatiotemporal features can be used for spatiotemporal prediction. Furthermore, considering multiple features, our method may apply to multivariate predictions. For instance, in the case of larceny count prediction by district, by considering the spatial features extracted from not only the larceny data but also robbery, burglary, and other crime data, the subsequent GBDT can estimate common/uncommon map patterns across crime types and might improve its predictive performance.

However, the properties of the proposed method remain largely unexplored. For example, it is unclear whether it works when the source and target areas are heterogeneous, such that the former are urban areas, but the latter is a mountainous area. It is also necessary to investigate whether a larger number of source areas yields better predictive accuracy in the target area.

The application of our method to a wide variety of urban and regional problems, such as environmental monitoring and COVID-19 spread, would also be an important next step. These applications would be valuable for improving predictive accuracy and analyzing common and uncommon tendencies across cities/regions.